\documentclass{emulateapj}

\newcommand{\kep}{{\it Kepler}}
\newcommand{\Kepler}{{\it Kepler}}
\newcommand{\Spitzer}{{\it Spitzer}}

\usepackage{hyperref}
\usepackage{breakurl}
\usepackage{amsmath}

\slugcomment{}

\shorttitle{K2 Photometry}
\shortauthors{Vanderburg \& Johnson}

\begin{document}


\title{A Technique for Extracting Highly Precise Photometry\\ for the Two--Wheeled \kep\ Mission}


\author{Andrew Vanderburg\altaffilmark{1,2} \& John Asher Johnson\altaffilmark{3}}
\affil{Harvard--Smithsonian Center for Astrophysics, 60 Garden St., Cambridge, MA 02138}

\altaffiltext{1}{avanderburg@cfa.harvard.edu}
\altaffiltext{2}{NSF Graduate Research Fellow}
\altaffiltext{3}{David \& Lucile Packard Fellow}


\begin{abstract}
The original \kep\ mission achieved high photometric precision thanks to ultra--stable pointing enabled by use of four reaction wheels. The loss of two of these reaction wheels reduced the telescope's ability to point precisely for extended periods of time, and as a result, the photometric precision has suffered. We present a technique for generating photometric light curves from pixel-level data obtained with the two-wheeled extended \Kepler\ mission, K2. Our photometric technique accounts for the non--uniform pixel response function of the \kep\ detectors by correlating flux measurements with the spacecraft's pointing and removing the dependence. When we apply our technique to the ensemble of stars observed during the \Kepler\ Two--Wheel Concept Engineering Test, we find improvements over raw K2 photometry by factors of 2-5, with noise properties qualitatively similar to \Kepler\ targets at the same magnitudes. We find evidence that the improvement in photometric precision depends on each target's position in the \Kepler\ field of view, with worst precision near the edges of the field. Overall, this technique restores the median attainable photometric precision to within a factor of two of the original \Kepler\ photometric precision for targets ranging from 10$^{th}$ to 15$^{th}$ magnitude in the \Kepler\ bandpass, peaking with a median precision within 35\% that of \Kepler\ for stars between 12$^{th}$ and 13$^{th}$ magnitude in the \Kepler\ bandpass. 
\end{abstract}


\section{Introduction}

Since its launch in 2009, the \kep\ spacecraft has led to the discovery of thousands of transiting exoplanet candidates, including the first multi-transiting exoplanet system \citep{kepler9}, the first Earth sized exoplanets \citep{kepler20}, the first Mars sized exoplanet \citep{koi961}, the first Mercury sized exoplanet \citep{barclay}, the first Earth-sized planet in the habitable zone of its host star \citep{quintana}, and the first circumbinary planets \citep{kepler16}. \Kepler's large exoplanet sample has led to statistical analyses of the frequency of exoplanets around solar mass stars \citep{howard, youdin, fressin}, around cool stars \citep{swift, mortonswift}, and in their stars' habitable zones \citep{dressing, petigura, foremanmackey}. \Kepler's scientific contributions extend beyond exoplanets to fields ranging from asteroseismology \citep{asteroseimology} to active galactic nuclei variability \citep{agn}.

\kep's scientific impact comes from its ability to monitor a large field of view with highly precise, high-duty cycle photometry for extended periods of time \citep[$\simeq$ 10 parts per million (ppm) per 6--hours for 10$^{th}$ magnitude stars,][]{christiansen}. This ability was compromised in May 2013, when the second of four reaction wheels on the \Kepler\ spacecraft failed, leaving the telescope unable to point precisely at its original target field. Since then, a new \Kepler\ mission concept, named K2, has been planned and executed, in which the spacecraft balances itself against Solar radiation pressure by pointing along the plane of the spacecraft's orbit and using thrusters to mitigate the residual 
spacecraft drift \citep{howell}. In this configuration, \Kepler\ is able to conduct science operations for approximately $\sim 75$ days at a time with decreased photometric precision due to decreased fine--pointing control, before moving onto a new field to observe. Initial tests of this observing strategy described in \citet{howell} indicate that raw K2 aperture photometry is a factor of 3 to 4 less precise than photometry from the original \Kepler\ mission. 

However, unstable pointing is not the death--knell for precise photometry. Since the discovery of transiting exoplanets, other space telescopes without the stable pointing of the original \Kepler\ Mission have been used to make precise photometric observations. Astronomers have developed techniques to reduce data from space telescopes, in particular \Spitzer, that substantially improved photometric precision by correcting for artifacts caused by the motion of the spacecraft \citep{knutson, ballardspitzer,blissmap}. These techniques have enabled telescopes like \Spitzer\ to become valuable resources for the study of transiting exoplanets. 

In this paper, we present a technique for extracting and correcting K2 photometry similar to those developed for \Spitzer\, but optimized for the peculiarities of the K2 mission. In Section \ref{datareduction}, we describe the procedure, and in Section \ref{results}, we assess the photometric precision of K2 using our reduction technique on the ensemble of stars observed during the \Kepler\ Two--Wheel Concept Engineering Test, conducted in February 2014. In Section \ref{discussion}, we discuss the performance of our technique and the implications for K2 science.

\section{K2 Data Reduction}\label{datareduction}
\subsection{Aperture Photometry}\label{apertures}

\begin{figure*}
\epsscale{1}
  \begin{center}
      \leavevmode
\plotone{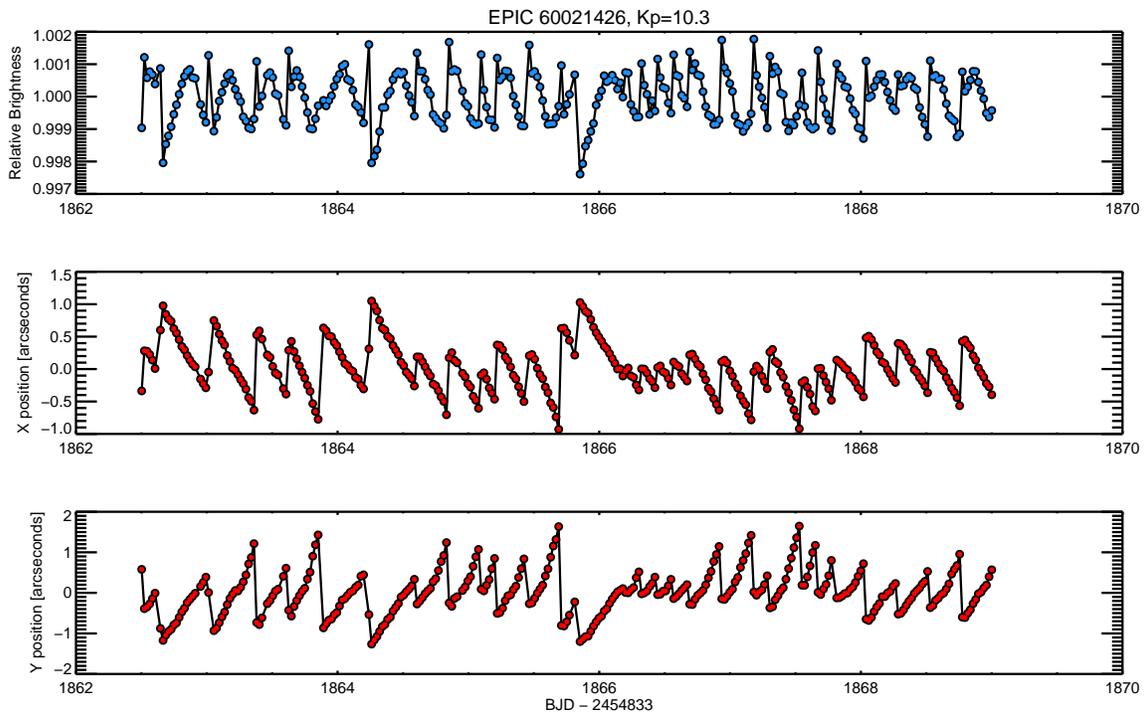}
\caption{Top: Raw K2 light curve (with low frequency variations removed). Middle: Horizontal centroid position versus time. Bottom: Vertical centroid position versus time. The light curve is dominated by noise corresponding to the changes in the spacecraft's pointing. The centroid measurements shown here were made by fitting a two dimensional Gaussian to the image of the star. The first 2.5 days of data have been excluded from this plot, as well as all data flagged by the \Kepler\ pipeline as having poor quality. For this particular star, the correlations between flux and X and Y positions are roughly of the same magnitude, but depending on the position on the detector, this is not generally true.} \label{rawlc}
\end{center}
\end{figure*}

Unlike \Kepler\ data, K2 data is dominated by noise due to the spacecraft's pointing jitter. Our approach is to extract aperture photometry and image centroid position information (as a proxy for the spacecraft's motion) from the K2 pixel--level data and correct the photometry using our knowledge of the spacecraft's pointing.

\Kepler\ collected the data we use in our analysis during a 9 day test of the new two--wheeled operation mode conducted between 4 February 2014 and 13 February 2014. During those nine days, \Kepler\ pointed to a field centered at RA~$ = -1.35$~degrees, DEC~$ = -2.15$~degrees and monitored about 2000 targets with 30 minute ''long cadence'' exposures and 17 targets with 1 minute ''short cadence'' exposures. After the first 2.5 days of the test, \Kepler\ achieved fine guiding control, that is, the spacecraft locked onto the center of its field, its pointing stabilized, and \Kepler\ began collecting high quality data. We focus primarily on the long cadence data collected after \Kepler\ achieved fine guiding control.

We downloaded target pixel files for all K2 engineering targets from the Mikulski Archive for Space Telescopes (MAST) and measured raw photometry for each star in the dataset. Due to spacecraft datalink bandwidth limitations, \Kepler\ does not download a full--frame image for every photometric timestep. Instead, small sub--images, or ``postage stamps,'' are available for each target. Due to the increased uncertainty in the spacecraft's pointing, the K2 Engineering Test postage stamps are 50 by 50 pixel squares, much larger than the \Kepler\ postage stamps. For each of these postage stamps, we start by defining an aperture mask in two ways: extracting a circle of pixels around the target star, and extracting a region defined by flux from the telescope's pixel response function (PRF). We analyze the photometry from each method separately and choose the one that provides the best photometric precision as measured using the procedure outlined in Section \ref{precision}. All of our apertures are stationary; the movement of the star on the \Kepler\ detector is small enough (typically within one pixel) that moving apertures are not required. The first mask we define is a circular aperture centered on the target star, with a radius determined by the \Kepler\ bandpass magnitude ($K_p$) of the target star. After optimizing the mask radius for robustness by trial and error, we arrived at a mask radius function that varies between 13 pixels for the brightest targets ($K_p = 9$) to 6 pixels for stars fainter than $K_p = 13.5$. The circular masks are intentionally large and conservative in order to prevent flux from spilling out of the aperture as the star drifts. The large circular masks typically give better photometric precision for saturated stars and stars with close companions.

In addition to defining large circular apertures around the target stars, we also define more aggressively fitted apertures using the \Kepler\ PRF \citep{keplerprf}. We fit the PRF downloaded from the MAST to one long--cadence image of the target star, using a Levenberg--Marquardt least squares minimization algorithm \citep{mpfit}. We fit for four free parameters: horizontal and vertical centroid positions, an amplitude, and a background offset. For stars with $K_p < 13$, we define our aperture mask as those pixels for which the PRF model flux is greater than 0.005\% of the total PRF model flux. For stars with $K_p > 13$, we define our aperture mask as pixels where the PRF model flux is greater than 0.1\% the total PRF model flux. The flux levels that define the apertures were determined through trial and error to maximize photometric precision. This procedure yields a tightly shaped aperture about the stellar image on the detector. The smaller apertures limit background light contamination and thus improve photometric precision for faint, background--limited stars, while decreasing the robustness of aperture photometry for bright stars, particularly for saturated stars. 

After defining the apertures, which remain stationary over the length of the K2 observations, we extract photometry for each K2 image using the \texttt{FLUX} tag from the \Kepler\ FITS files data structure. First, for each image we estimate the background flux by taking the median value of the background pixels that lie outside of our aperture. We have experimented with estimating the background flux using a robust (outlier--rejected) mean calculation, but found no improvement over the median. We then subtract the background flux from the image and sum the pixels within the aperture. After extracting photometry for all K2 datapoints for a particular star, we divide by the median measured brightness to roughly continuum--normalize our photometry.

We then estimate the position of the star on the detector for each K2 datapoint using two methods. First, we calculate the centroid using a ``center of flux'' calculation given by

\begin{equation}
x_{c} = \frac{\sum\limits_{i} x_{i} f_{i}}{\sum\limits_{i} f_{i}},\ \ y_{c} = \frac{\sum\limits_{i} y_{i} f_{i}}{\sum\limits_{i} f_{i}},
\label{eqn:comcent}
\end{equation}

\noindent where $x_{c}$ and $y_{c}$ are the centroid positions in the horizontal and vertical dimensions of the K2 image, respectively;  $x_{i}$ and $y_{i}$ are the individual horizontal and vertical positions of pixels; and $f_{i}$ is the flux in each individual pixel. 

We also estimate the position of the star on the detector by fitting a multivariate Gaussian to the image with a Levenberg--Marquardt minimization routine. We fit a Gaussian model to the star instead of the \Kepler\ PRF for computational speed. We allow the amplitude of the Gaussian, an offset, widths in the horizontal and vertical directions, a cross term that rotates an elongated Gaussian, and horizontal and vertical positions to vary. Specifically, the model is given by:


\begin{equation}
F(x, y) = A \exp{\left[-z - B(x - x_{c})(y - y_{c})\right]} + D,
\label{eqn:gausscent}
\end{equation}

\noindent where

\begin{equation}
z  = \frac{(x - x_{c})^{2}}{\sigma_{x}^2} +  \frac{(y - y_{c})^{2}}{\sigma_{y}^2},
\end{equation}

\noindent $A$ is the flux at the center of the star on the detector, $x$ and $y$ are the horizontal and vertical positions on the detector, respectively, $\sigma_{x}$ and $\sigma_{y}$ are the widths in the horizontal and vertical directions, respectively, $B$ is the amplitude of a cross term that rotates the elongated PSF along an arbitrary axis, and $D$ is an offset to account for the background.

For each K2 target, we automatically choose between the ``center of flux'' and Gaussian centroids based on which exhibit smaller root mean square (rms) residuals when fitted by a fifth--order polynomial. In most cases, the Gaussian centroids (Eqn~\ref{eqn:gausscent}) are more precise than the ``center of flux'' centroids (Eqn~\ref{eqn:comcent}). But in some cases, particularly for saturated stars exhibiting bleed trails, a Gaussian model does not describe the shape of the stellar image and the ``center of flux'' measurements are better estimators of the centroid position.

Extracting raw photometry and centroid positions from the K2 images results in flux and position time series characterized by jagged features occurring several times a day. Figure \ref{rawlc} shows an example of a ``raw'' light curve and image centroid positions for a 10$^{th}$ magnitude K2 target. The position of the star on the detector drifts on time scales of hours and then is quickly corrected back to its starting point when the spacecraft thrusters fire. The changes in position introduce significant artifacts into the raw K2 photometry, which must be corrected in order to achieve the photometric precision characteristic of the original \Kepler\ Mission. 

\subsection{Data Exclusion} \label{exclusion}

After extracting raw photometry, we begin by excluding certain data from further analysis. In particular, we exclude data from the first 2.5 days of the Two--Wheel Concept Engineering Test, which were taken before the spacecraft achieved adequate fine pointing control. These data exhibit larger variations in the image centroid position, and the photometric precision is worse than after fine guiding was established. We also exclude points labeled by the \Kepler\ pipeline as having poor quality, where the \texttt{QUALITY} tag of the \Kepler\ FITS file data structure was not 0. This cut excludes data with a variety of anomalies, including cosmic rays, pointing adjustments, and detector glitches. Typically, between 2\% and 4\% of datapoints are excluded at this stage.

\begin{figure}
\epsscale{1}
  \begin{center}
      \leavevmode
\plotone{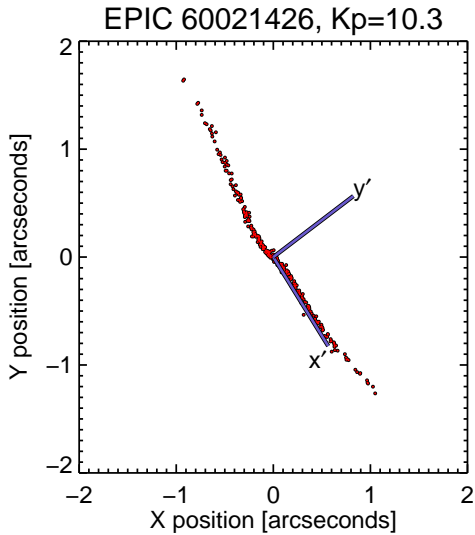}
\caption{Measured K2 centroid positions. The red dots are centroid positions measured by fitting a 2d Gaussian to the star on the detector. The spacecraft repeatedly moves back and forth along the same path on the detector. The wiggle near (0,0) is likely an artifact caused by estimating the \Kepler\ PRF as a two dimensional Gaussian. The purple lines are the axes of the rotated coordinate system, in which the $x'$ axis lies along the most significant eigenvector. The plot region is scaled to be the size of one \Kepler\ pixel. } \label{centroids}
\end{center}
\end{figure}

Finally, we exclude points taken while the spacecraft's thrusters were firing. The behavior of the \Kepler\ spacecraft during K2 data collection is to acquire the field using thrusters and stabilize the spacecraft's pointing with \Kepler's two remaining reaction wheels, while balancing the spacecraft against solar radiant flux in an unstable equilibrium. After approximately six hours of drifting off equilibrium due to the Solar wind, \Kepler\ fires its thrusters to bring the spacecraft closer to the equilibrium point. The result of this is that the spacecraft drifts only along its roll axis, and the movement of the spacecraft is one dimensional in an arc about \Kepler's boresight, or the center of its field. We observed that during thruster fires K2 photometry exhibits discontinuous behavior. Excluding these points yields improved photometric precision. 

We identify the points during which \Kepler's thrusters were firing through analysis of the image centroids. Because the spacecraft moves almost solely in the roll direction, the position of the spacecraft follows a one--dimensional curve on the detector to good approximation. We automatically detect the direction of motion using a procedure similar to that of a Principal Component Analysis (PCA). We take the image centroids measured both from the center--of--flux analysis and the Gaussian fits and separately calculate the covariance matrix between horizontal and vertical centroid positions and its eigenvectors. Because the eigenvectors of the covariance matrix are the basis in which the covariance matrix is diagonal (that is, the covariance is zero), the eigenvectors lie parallel and perpendicular to the direction of motion of the star. Figure \ref{centroids} shows an example of the path of the star on the detector, as well as the measured direction of movement. 

We then rotate the image centroids into new coordinates with the $x^\prime$ axis along the direction of the strongest correlation(that is, the eigenvector associated with the largest eigenvalue). We fit a fifth--order polynomial to both sets of the transformed centroids, while excluding 5--$\sigma$ outliers from the fit. Artifacts from the centroiding algorithm (like the wiggle seen in Figure \ref{centroids}) complicate the star's measured path along the detector to the point where a high order polynomial fit is required. At this point, we choose between the center--of--flux centroids and the Gaussian--fit centroids and proceed with the set of centroid measurements that provide the lower rms residuals to the polynomial fit. 

We calculate the arclength along the polynomial curve, where arclength, $s$, is defined as:

\begin{equation}
\label{eqn:arclength}
s= \int_{x'_0}^{x'_1}\sqrt{1+\left( \frac{dy'_p}{dx'}\right)^2} dx'
\end{equation}

\noindent where $x^\prime_0$ is the transformed $x$ coordinate of the point with the smallest $x^\prime$ position, and $y^\prime_p$ is the best--fit polynomial function. We then differentiate arclength with respect to time to calculate $ds/dt$. When \Kepler's thrusters fire there are sharp changes in the position of the spacecraft, so $ds/dt$ is significantly different from the typical drifting behavior. We identify the points during which the spacecraft fires its thrusters as 5--$\sigma$ outliers from the distribution of $ds/dt$ points and exclude them from further calculations.

\subsection{Self--Flat--Fielding Correction}
\begin{figure}
\epsscale{1}
  \begin{center}
      \leavevmode
\plotone{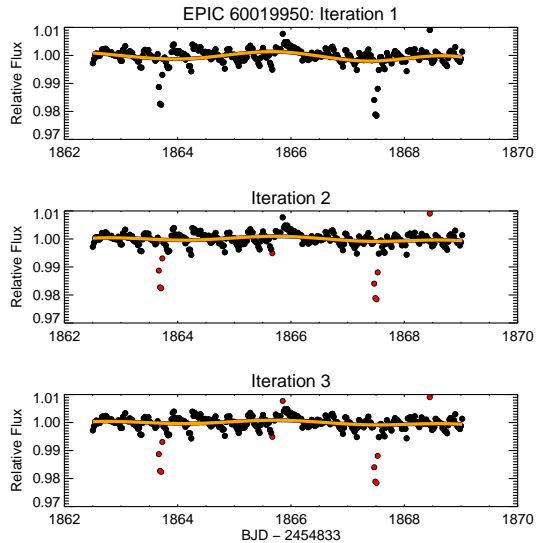}
\caption{Example of the iterative low--frequency B-spline fitting routine. Top: The raw light curve of an eclipsing binary star from \citet{k2eb} fit with a B-spline (the solid orange line). No points are excluded so the spline fit is pulled towards outlier points. Middle: After the first iteration, several points (shown in red) are excluded, and the spline fit is improved. Bottom: After one more point is excluded, the fit converged and the spline is an accurate estimate of the low--frequency modulations.} \label{lowfreq}
\end{center}
\end{figure}
After excluding points with poor photometric performance, we used a ``self--flat--fielding'' (SFF) approach to remove photometric variability caused by the motion of the spacecraft. We start by isolating the short--period (timescales $\tau \lesssim 24$~hour) variability of the light curve, which we assume  is dominated by the 6--hour spacecraft pointing jitter for most stars. We estimate the low frequency variability in the light curve by iteratively fitting a basis spline (B-spline) with breakpoints every 1.5 days in the light curve. We iteratively fit the B-spline, identify and exclude 3--$\sigma$ outlier points from the fit, and re-fit the B-spline, repeating this process until convergence (typically less than five iterations). This process is illustrated in Figure \ref{lowfreq}. We found this approach to be a good balance between the desire to retain enough flexibility needed to describe the stellar variation and avoiding over--fitting to noise caused by pointing jitter.  We then isolate high frequencies by dividing our raw aperture photometric time series by the final B-spline fit.

\begin{figure}
\epsscale{1}
  \begin{center}
      \leavevmode
\plotone{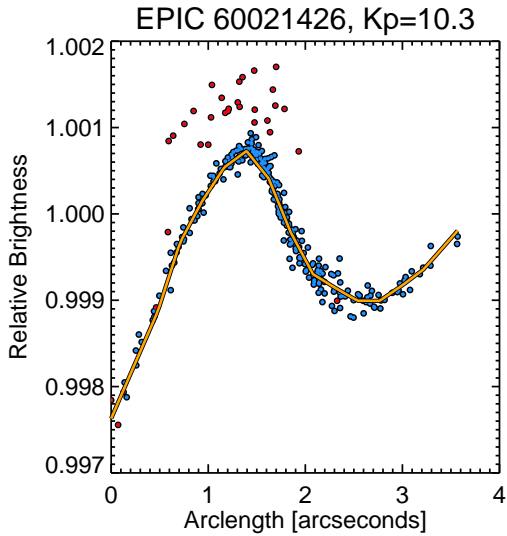}
\caption{High--pass filtered flux as a function of centroid position, parameterized by arclength, $s$. Blue points are flux measurements taken while the spacecraft is drifting. Red points mark data taken during thruster fires. These points are clear outliers and are therefore excluded. We model the dependence of measured flux on centroid position with a piecewise linear interpolation, shown in orange.} \label{fvsarclength}
\end{center}
\end{figure}

The short--timescale variations remaining in the light curve are a combination of astrophysical variability (such as eclipse/transit events, flares, oscillations or flicker) and noise due to the spacecraft's pointing jitter, whether caused by uneven pixel sensitivity \citep[e.g.][]{ballardspitzer} or different amounts of flux falling outside of the aperture \citep[e.g.][]{blissmap}. We separate astrophysical variability from pointing jitter by removing the noise component that correlates with the position of the star on the detector, as parametrized by arclength, $s$ (Eqn~\ref{eqn:arclength}). Figure~\ref{fvsarclength} shows a plot of the high--pass filtered  photometry versus the arclength. The dependence of measured flux on the image position is evident.

We divide the data points into 15 bins in arclength and calculate the mean within each bin with $3\sigma$ outlier exclusion. The outlier exclusion serves to reduce the influence of short--timescale astrophysical noise on the correction. We then perform a linear interpolation between the mean of each bin. The result is a ``correction'' that can be applied to raw aperture photometry to remove the effects of pointing jitter. We chose a piecewise linear function to model the flux variation because it was simpler than other alternatives (like a spline or polynomial fit), and produced light curves of the same quality. Finally, we applied the correction by dividing the raw aperture photometry time series by the piecewise linear fit. The resulting light curve gives a much better estimate of the low--frequency variations than the raw aperture photometry, so we recalculated the low--frequency component with an iterative B-spline fit to the corrected lightcurve, which we removed from the uncorrected lightcurve, and used to recalculate the SFF correction. This process typically converged after one or two iterations.

We remove variation due to centroid position changes in only one dimension, specifically along the arclength of the polynomial fit. The spacecraft's pointing was repeatable enough over the 6.5 days of the fine pointing test to not require removal of variation transverse to the direction of drift on the detector. However, our technique is easily generalizable to removing photometric artifacts due to image centroid variations in two dimensions. This might be important for future K2 data releases for longer campaigns, particularly if the boresight of the spacecraft drifts over the course of 75 days of observations, and the path of the spacecraft's pointing jitter cannot be approximated as one--dimensional. 

\begin{figure*}
\epsscale{1}
  \begin{center}
      \leavevmode
\plotone{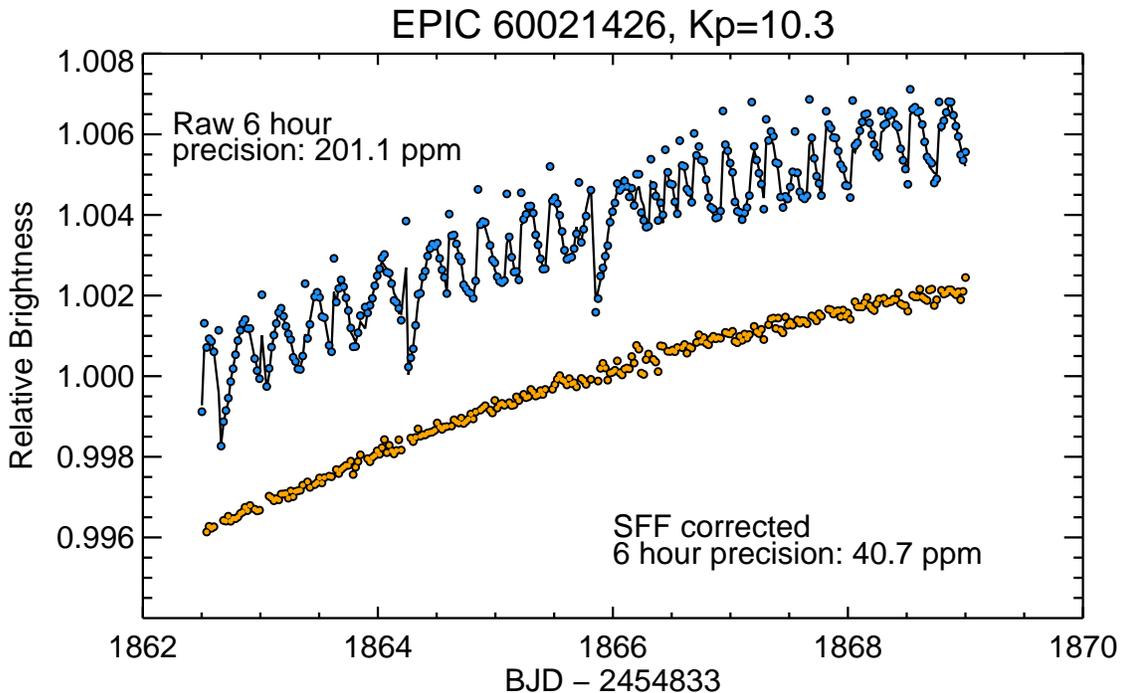}
\caption{Comparison between raw K2 photometry and SFF corrected photometry. Blue points are the raw K2 light curve, which is vertically offset for clarity. Orange points are the SFF corrected light curve, which shows a substantial improvement in photometric precision. The black line underneath the raw data is the SFF model. The photometric precision of this star is slightly worse than the median precision we achieve for 10$^{th}$ magnitude stars. The SFF technique preserves astrophysical signals like transits (see Section \ref{injectionsection}) and starspot modulation, which is evident in this light curve. This bodes well for the prospects of detecting stellar rotation periods with K2.} \label{finalcorrected}
\end{center}
\end{figure*}

Figure \ref{finalcorrected} shows the result of the SFF correction and data exclusion on the K2 photometry. For the target star EPIC\,60021462, the SFF correction decreases the scatter measured on 6--hour timescales by a factor of 5. 

\section{Results}\label{results}
\subsection{Photometric Precision}\label{precision}

\begin{figure*}
\epsscale{1}
  \begin{center}
      \leavevmode
\plotone{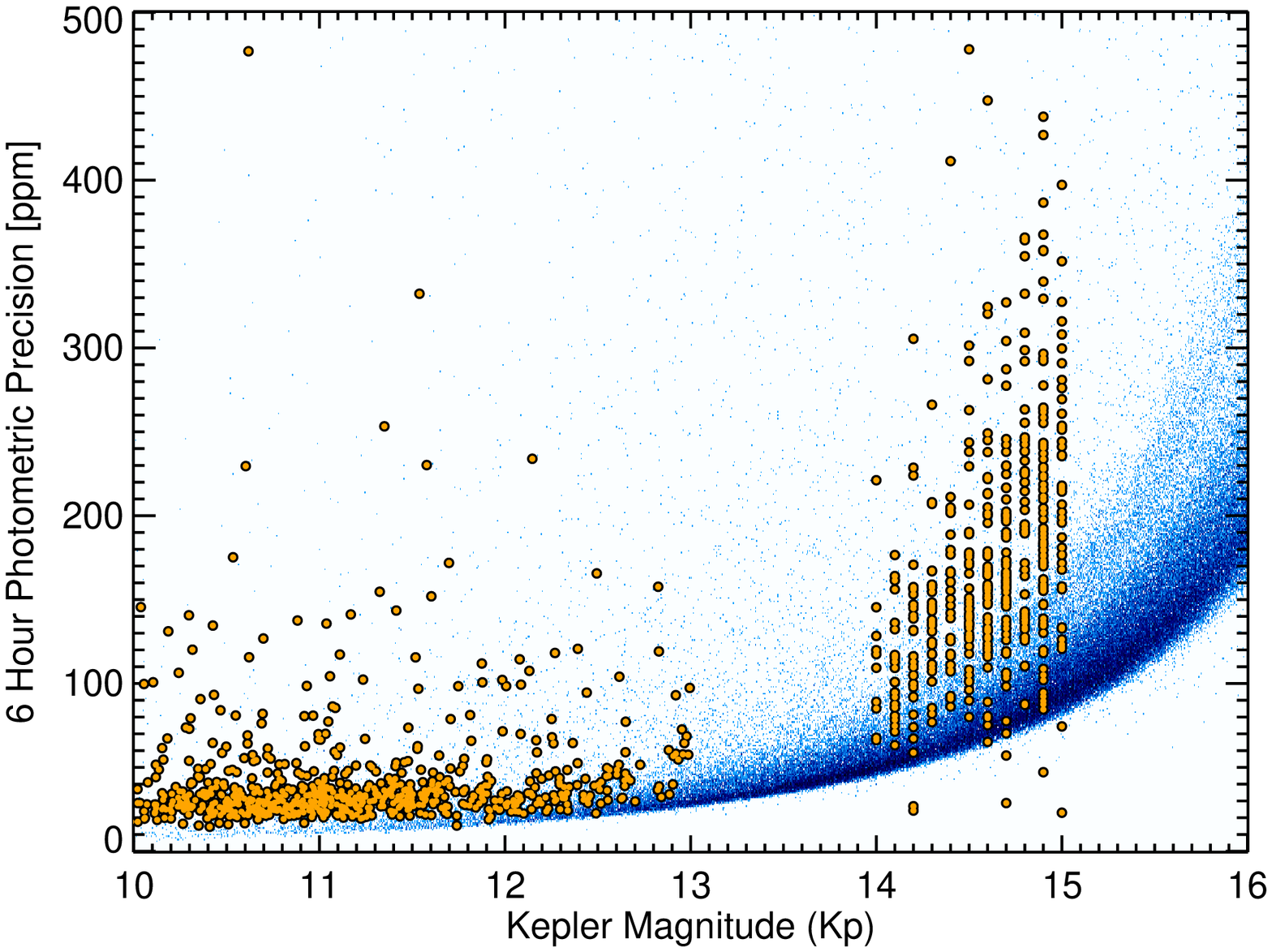}
\caption{Photometric precision of K2 versus that of \Kepler. Orange points are stars observed by K2 during the Two--Wheel Concept Engineering Test, and blue indicates a density of stars observed by \Kepler\ during Quarter 10 of its operation. K2 photometry is consistently less precise than \Kepler's, but for some stars approaches its precision. The gap between $K_p = 13$ and $K_p = 14$ is due to magnitude cuts in the selection of dwarf star targets for the Engineering Test.} \label{cdppvskp}
\end{center}
\end{figure*}

We applied our SFF reduction technique to all targets observed during the K2 Two--Wheel Concept Engineering Test to assess its performance. The total runtime was about 5 hours on a laptop computer. We find that for almost all targets, our reduction improves the photometric precision over raw aperture photometry. Exceptions to this rule include stars with rapid astrophysical variation such as Cepheid variable stars, contact binaries and giant stars with large--amplitude oscillation modes. 

We assess the quality of our K2 light curves by estimating their photometric precision over a six--hour window, a metric similar to the Combined Differential Photometric Precision, CDPP, metric used by the \Kepler\ pipeline. We perform this measurement using a method inspired by the \Kepler\ Guest Observer tools routine \texttt{kepstddev}\footnote{\url{http://keplerscience.arc.nasa.gov/ContributedSoftwareKepstddev.shtml}}. For each target, we calculate the standard deviation within a running bin of 13 long--cadence measurements in length, divide by $\sqrt{13}$ and report the median value of the running standard deviation as the target's photometric precision. For each star, we calculate the photometric precision of the light curves produced from both the large circular apertures and the small PRF--defined apertures, as described in Section \ref{apertures} and retain the smaller of the two estimates.

We find that photometric precision of the K2 sample compares favorably to the photometric precision of the \Kepler\ mission. To ensure a differential comparison, we downloaded light curves for all \Kepler\ targets observed in Quarter 10 from the MAST and estimate photometric precision of the \texttt{PDCSAP\_FLUX} light curve using the same procedure as we used for the K2 engineering data. Figure \ref{cdppvskp} shows the measured photometric precision for dwarf stars observed by K2 and \Kepler\ as a function of \Kepler\ magnitude. We select the dwarfs observed by \Kepler\ by taking all stars with $\log{g} \geq 4$ according to the \Kepler\ Input Catalog \citep[KIC;][]{kic}, and we selected dwarfs observed by K2 as those stars whose target list was designated as \texttt{cool\_star} or \texttt{GKM\_dwarf} according to the K2 engineering test target list\footnote{\url{http://archive.stsci.edu/missions/k2/tpf_eng/K2_E2_targets_lc.csv}}. We find that in the best cases, the photometric precision of K2 approaches that of \Kepler, but the typical precision of a K2 target is consistently worse by roughly a factor of 1.3--2. We also note that there is more scatter in the photometric precision of K2 targets than in \Kepler\ targets, and investigate this further in Section \ref{positiondependent}.

We summarize the results of our SFF photometric analysis in comparison to \Kepler's original precision in Table \ref{precisioncomp}. Raw K2 photometry exhibits worse photometric precision than the original \Kepler\ mission by at least a factor of four, and is comparatively worse for brighter target stars. The SFF technique presented in this paper restores K2 photometry to within a factor of two of the precision of \Kepler\ photometry when comparing the median photometric precision of each data set.

\begin{table}[t!]
\begin{center}

\caption{Median Photometric Precision}
\begin{tabular}{llll}
\hline\hline
$K_{p}$ & Raw K2 & SFF K2 & \Kepler\ \\
\hline
10-11            & 170    & 31   & 18         \\
11-12            & 163    & 33   & 22         \\
12-13            & 157    & 40   & 30         \\
14-15            & 365    & 164  & 81     \\
\hline\hline
\end{tabular}
\end{center}
\tablecomments{These photometric precision measurements represent the median precision of all dwarf stars observed by K2 and \Kepler\ and are reported in parts per million. \Kepler's photometric precision specifically refers to \texttt{PDCSAP\_FLUX}.}
\label{precisioncomp}

\end{table}

\subsection{Position Dependent Photometric Precision}\label{positiondependent}

\begin{figure}
\epsscale{1}
  \begin{center}
      \leavevmode
\plotone{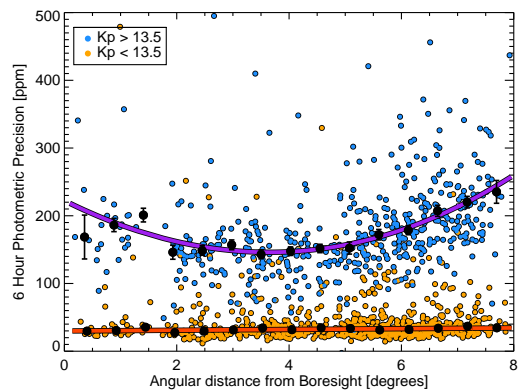}
\caption{Photometric precision of K2 versus distance from the spacecraft's boresight. Bright stars are shown in orange, and faint stars are shown in blue. Large black dots indicate the median precision of the 15 radial bins, and the thick lines represent the best--fit polynomials to the medians. Error bars on the binned points reflect the scatter within the bin, and in many cases, are smaller than the size of the point. The stars' photometric precision has been corrected to remove effects due to the brightness of the star within the bin. A quadratic dependence of photometric precision on the star's boresight is evident for faint stars, and a weak linear dependence is measured for bright stars, with worsening precision at larger distances.} \label{cdppvsdegree}
\end{center}
\end{figure}

Figure \ref{cdppvskp} shows evidence for large scatter in the photometric precision of K2 targets at a given magnitude, so we investigated factors beyond the brightness of a target that affect its photometric precision. We found a correlation between a star's position on the \Kepler\ detector and its measured photometric precision. We calculated the angular distance of each star from the center of the \Kepler\ boresight, held during the Engineering Test at RA~$ = -1.3507626$~degrees, DEC~$ = -2.1523890$~degrees (T. Barclay 2014, private communication). Figure \ref{cdppvsdegree} shows our measured 6--hour photometric precision of the K2 data as a function of angular distance of the target from the center of the boresight. 

We divided the K2 engineering sample of bright stars into 15 bins in radial distance and found the median of each bin. For the bright stars ($K_p < 13$), we noticed a slight linear trend of worsening photometric precision as a function of angular distance from the boresight. We fit a line to the medians of the radial bins (with errors estimated from the scatter within each bin) using a Markov Chain Monte Carlo algorithm with an affine invariant ensemble sampler \citep[adapted for IDL from the algorithm of][]{goodman, emcee}. Any polynomial model more complex than a line resulted in high order polynomial terms consistent with zero and was not justified by the Bayesian Information Criterion \citep[BIC, ][]{bic}, a test which balances model complexity with improvements in the fit. We measured a slight radial increase in 6--hour photometric scatter of $0.57 \pm 0.15$~ppm/degree. Our best--fit polynomial was:

\begin{equation}
\label{eqn:brightline}
P_{6,bright} = 29.8 + 0.57 r
\end{equation}

\noindent where $P_{6,bright}$ is the 6 hour photometric precision for bright stars in ppm and $r$ is the distance from the boresight in degrees. The worsening photometric precision as a function of distance from the boresight is likely because stars farther from the center of the field move a larger distance on the detector as the spacecraft's roll drifts. This introduces larger and more complex photometric variations farther from the center of the spacecraft's field of view, which are more difficult to correct and result in larger photometric scatter.

The median photometric precision of faint stars ($K_p \gtrsim 14$) also shows a dependence on the angular distance from the center of the boresight. We investigated this dependence by dividing the K2 sample of faint stars into 15 bins in radial distance and noticed that the trend in radius was non--monotonic--- photometric scatter increases both very close to the boresight and very far away. We fit a second order polynomial to the median values of the radial bins (using the same technique and model complexity tests as before) and found that photometric precision is best at a distance of $3.62 \pm 0.11$ degrees from the boresight. Our best--fit polynomial was:

\begin{equation}
\label{eqn:faintcurve}
P_{6,faint} = 220 - 43 r + 5.9 r^2
\end{equation}

\noindent where $P_{6,faint}$ is the 6 hour photometric precision for bright stars in ppm and $r$ is the distance from the boresight in degrees. A similar non-monotonic dependence of precision on distance from the boresight was noticed by \citet{christiansen}, who attributed the effect to the quality of the spacecraft's focus, which is best between 3-6 degrees from the center of the field. We speculate that the focus quality affects faint stars more than bright stars because they are closer to background--limited photometric precision. The degraded pointing stability of the two--wheeled \Kepler\ spacecraft forces the use of larger photometric apertures, which increases the amount of scattered background light in the aperture, adding to photometric uncertainties.

\subsection{K2 Noise Power Spectrum}\label{noisepowerspectrum}

The SFF correction primarily removes red noise (that is, low--frequency noise) introduced to K2 data by pointing drift and correctional thruster fires. We investigated the noise power spectrum of K2 data, in comparison to that of \Kepler\ data, to understand the extent to which red noise is introduced into K2 data. We selected two stars observed by K2, the 10$^{th}$ magnitude EPIC\,60021426 and the 14$^{th}$ magnitude EPIC\,60029819, with photometric precisions close to the median for K2 targets of their brightness and after removing long--period stellar variability we calculated their power spectra via Fast Fourier Transform (FFT). We then selected two stars observed by \Kepler\ of similar brightnesses, KIC\,009579208 ($K_{p}$ = 10.3) and KIC\,012066509 ($K_{p}$=14.7) with photometric precisions close to the median for \Kepler\ targets of their brightness. We examined a baseline of photometry of exactly the same length as that of the K2 data and calculated the noise power spectra for the \Kepler\ targets in the same way. Inspection of the power spectra showed no significant differences between the noise properties of \Kepler\ and K2 targets, other than that the K2 data has consistently higher levels of noise at all frequencies. Over the short time baseline we tested, we do not see obvious excesses of noise at low frequencies corresponding to the interval between K2 thruster fires.

\subsection{Transit Injection Tests}\label{injectionsection}

We tested the ability of our reduction technique to recover exoplanet transits by injecting artificial signals directly into the K2 pixel--level data and passing the modified data through our reduction pipeline. Because our technique uses the star's flux as a flat--field, it is possible that flux decrements, like transit events, could bias the flux correction and suppress or distort transit events.  We calculated a model transit light curve based on transit parameters from the NASA Exoplanet Archive \citep[][]{nea}, using a \citet{mandelagol} model, as implemented by \citet{exofast}, with limb darkening parameters from \citet{claret}. We then scaled our PRF model from Section \ref{apertures} by the flux decrement from the transit light curve model and subtracted this scaled PRF from each K2 image. After the injection, we performed the data reduction in exactly the same manner as described in Section~\ref{datareduction}.

\begin{figure*}
\epsscale{1}
  \begin{center}
      \leavevmode
\plotone{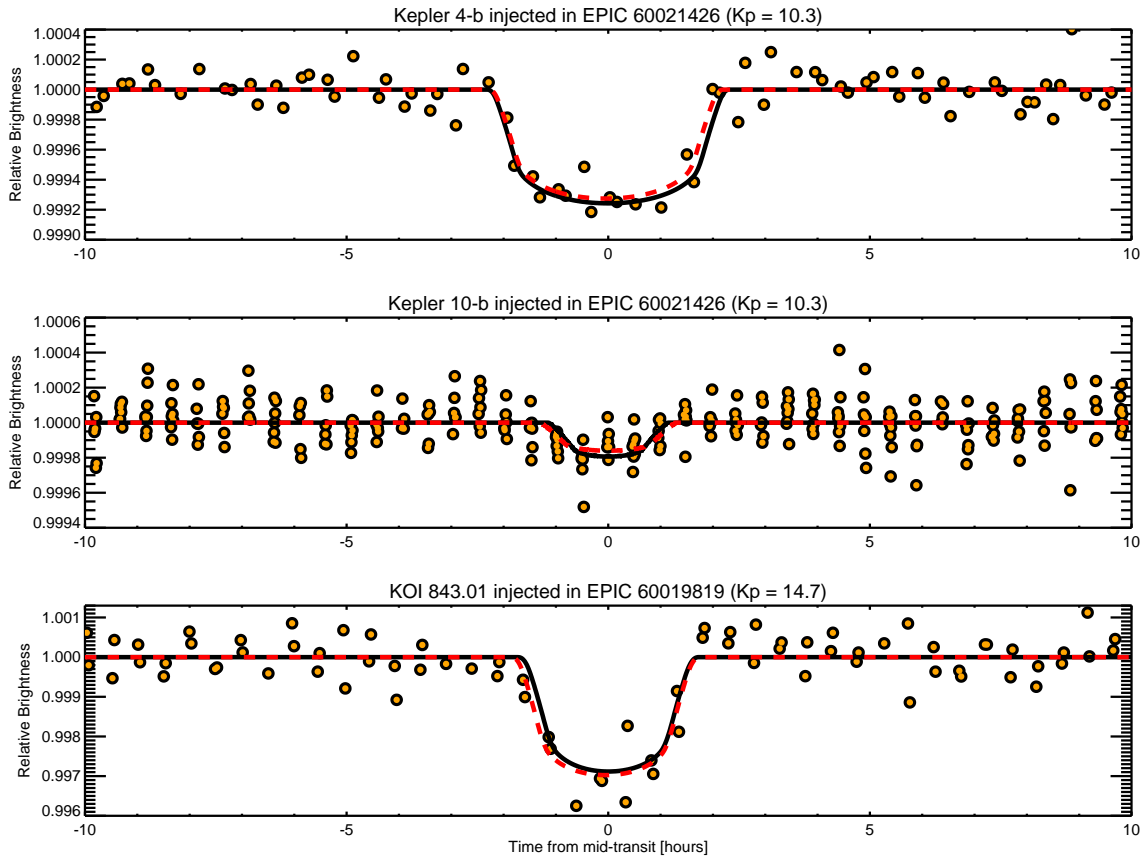}
\caption{Transit signals injected into raw K2 data and recovered with the SFF technique. The black solid line is the model injected into the pixel level data, the dashed red line is the best--fit model, and the orange dots are the recovered K2 photometry. Top: Two transits of Neptune sized Kepler 4-b injected into a typical bright K2 target. Middle: Eight transits of rocky super--Earth Kepler 10-b injected into a typical bright K2 target. Bottom: Two transits of Saturn sized KOI 843.01 injected into a typical faint K2 target. The close correspondence between the injected model and best--fit model demonstrates our ability to recover transits with minimal distortions to the light curves.} \label{injections}
\end{center}
\end{figure*}

We injected transit models of several different exoplanets into two K2 target stars, one bright ($K_p = 10.3$) and one faint ($K_p = 14.7$), with 6--hour photometric precisions close to the median K2 target for their brightness. The recovered light curves are shown in Figure \ref{injections}, along with the injected signals. Visual inspection shows that the recovered light curves show minimal distortion of the transit signal. To quantify this, we fit the recovered light curves with a \citet{mandelagol} model \citep[once again as implemented by][]{exofast} and compared our recovered parameters  to our inputs. We fitted model transit light curves the data using a Markov Chain Monte Carlo algorithm with an affine invariant ensemble sampler while holding most parameters fixed. 

For the two deep transits (Kepler 4-b and KOI 843.01), we held all parameters fixed except for the orbital period, the ratio of the planet's radius to that of its star ($R_p/R_\star$) and the ratio of the stellar radius to the orbital semimajor axis ($a/R_\star$). Both of the injected signals had only two transit events, so fitting for the orbital period is equivalent to fitting for one of the two transit times, thus testing our ability to recover Transit Timing Variations (TTVs). For the shallow transit (Kepler 10-b), we froze the orbital period because of the smaller signal, but still fit for $a/R_\star$ and $R_p/R_\star$.  We found in each of the three cases shown in Figure \ref{injections}, the recovered best--fit $R_p/R_\star$, $a/R_\star$, and, where applicable, orbital periods were within the 1--$\sigma$ uncertainties of the input parameters. Evidently, at the level of precision attainable with 6.5 days of K2 data, our SFF technique does not distort or suppress transit light curves. That being said, it is possible that with more data, transit parameter determinations will be precise enough to notice a bias. In that case, it will be necessary to incorporate the SFF technique into light curve fitting, in order to ensure that the SFF corrections are not biased by transit events.

\section{Discussion}\label{discussion}

\subsection{Algorithm Performance}

Our technique is designed to be generally applicable to most or all K2 targets and to be fast enough to be run on a personal computer. Because the algorithm is meant to be generally applicable to the K2 sample, there is room for photometric improvement on individual stars by ``hand--tweaking'' at various steps in the algorithm. This process can improve the photometric precision of individual stars when objects of interest are revealed using the general--purpose pipeline. 

The ability to adjust the reduction algorithm by hand on individual interesting targets is enabled by the fact that the technique presented in this paper relies only on data collected from one single K2 postage stamp. On the other hand, the technique could presumably be made more efficient and reliable by using data from multiple, independent K2 apertures. In particular, the motion of the spacecraft can be estimated from any number of bright but unsaturated stars to higher precision than is possible for the fainter targets in the K2 sample. 

Finally, our technique was designed to be computationally efficient to run multiple times on the entire K2 dataset. As presently constructed, processing one K2 light curve takes  $\simeq$~10 seconds per 9 day light curve on a laptop computer. The majority of the computer time is spent fitting the \Kepler\ PRF to define the aperture and the Gaussian model to each image to estimate the centroid. It is possible that more computationally expensive operations, like extracting photometry by fitting the \Kepler\ PRF to each image, could yield better photometry but the computational expense would be high. 

\subsection{Prospects for Continuing \Kepler's Science}
\subsubsection{Exoplanet Detection}

Exoplanet detection was the primary goal of the original \Kepler\ mission and remains a driving scientific motivation for the extended two--wheeled \Kepler\ mission. One goal of the K2 mission is to study planets around bright stars, in particular those bright enough for radial velocity (RV) followup, as well as planets around low--mass stars and stars in open clusters \citep{howell}. The planets most amenable to RV followup are at least roughly Earth sized and have short periods, ($\lesssim$~5~days) in order to boost the RV semiamplitude that must be measured \citep{howard13, pepe13}. For this type of planet orbiting a G-type dwarf, the transit depth would be roughly 100~ppm and would transit $\simeq$ 20 times during a K2 observing campaign. With a nominal transit duration of 3 hours (shorter than the timescale of most of the red noise in bright K2 light curves), and assuming a median instrumental 6--hour precision of $\simeq$~40~ppm, an Earth--size planet with a 3 day period  should be detectable at roughly 7--$\sigma$ around a typical bright ($K_{p} \lesssim 13$) G-dwarf. 

Ultra--short--period planets \citep[e.g.][]{ultrashort} could be difficult to discover if the planet's orbital period is near the time scale of the \Kepler\ thruster fires, which control the telescope pointing. In such cases, an SFF approach could suppress the transit signal because transit events could cluster around the same spot on the detector. It would be beneficial to search for ultra-short period planets using data detrended by both an SFF approach as well as other techniques like PCA to remove common modes to all stars \citep[e.g.][]{petiguraterra}. It could also be difficult to detect planets which have orbital periods near integer multiples of the interval between K2 thruster fires, especially if the transit time is during the K2 thruster fires. However, the duration of K2 thruster fires is shorter than the duration of a \Kepler\ 30 minute long cadence exposure, so for planets with transit durations longer than roughly 30 minutes, this will likely not prevent detection.

The detection of exoplanets around low mass stars like mid--to--late M-dwarfs is important to the K2 mission because these stars have small stellar radii, and therefore small, rocky planets can cause relatively large transit depths. This effect is the reason many of the first systems of sub--Earth--sized exoplanets, such as the Kepler--42 system, were discovered around faint M--dwarfs despite the photometry having a significantly higher noise level than brighter \Kepler\ targets \citep{koi961}. Planets like those orbiting Kepler-42 would be detectable in a K2 observing campaign around 14$^{th}$-15$^{th}$ magnitude stars and likely stars even fainter. With K2's typical precision of 650~ppm per 30 minutes for 14-15$^{th}$ magnitude stars, Kepler--42\,d's 1300~ppm transit on a 1.8 day orbit would give an $\simeq 13\sigma$ detection.  

Exoplanets in clusters could prove more challenging than field exoplanets for K2 to observe due to crowding effects. K2 photometric apertures must be larger than \Kepler\ apertures due to the imprecise pointing of the spacecraft, which increases the number of nearby stars that could contribute background light to target stars. Moreover, we noticed that the stars with the worst photometric performance in our reduction were those with other close stars of similar brightness near the edge of the aperture. This problem will be more ubiquitous for observations of stars in clusters than those in the field, which could complicate data reduction. For stars with neighbors close to their photometric apertures, it may be necessary to extract photometry by simultaneously fitting the \Kepler\ PRF to multiple stars.

\subsubsection{Flicker}
One of the many unanticipated results from \Kepler\ was an observational correlation between short timescale photometric variability  in light curves, referred to as ``flicker,'' and the surface gravity of the star \citep{bastien, bastien14, kippingflicker}. Flicker is an indicator of stellar granulation, the amplitude of which is determined by the surface gravity of the star ($\log{g}$). Using data from the original \Kepler\ mission, \citet{bastien} could measure the flicker of G-type dwarf stars at levels of $\simeq 30$~ppm per 30--minute exposure in some cases. 

The increased noise in K2 data could limit the stars on which flicker measurements are possible. If the typical precision of K2 data is roughly 30--40~ppm per six hours, or 100--150 ppm per 30 minutes, it could be difficult to measure flicker (as presently defined) for stars with $\log{g} \gtrsim 3.5$, for which \citet{bastien} predict a flicker of $\simeq 100$~ppm. This would effectively preclude flicker and therefore $\log{g}$ measurements for dwarf stars with K2.

Making flicker measurements from K2 data for dwarf stars may be possible using a more specialized light curve extraction algorithm, or by using a more extreme filtering technique than was necessary for the pristine \Kepler\ data on which the flicker measurements were pioneered.

\section{Summary}

We have developed a technique to reduce K2 data using a ``self--flat--fielding'' (SFF) approach and have demonstrated its performance on the sample of targets from the \Kepler\ Two--Wheel Concept Engineering Test. We find that our technique restores the median precision of K2 data to within a factor of two of the precision of \Kepler\ data for stars with $10<K_{p}<15$ and to within 35\% the precision of \Kepler\ data for targets with $12 < K_p < 13$. In the best cases, the precision of K2 data approaches that of \Kepler\ data. The SFF approach improves raw K2 photometry by a factor of 2-5.  The SFF technique is able to recover injected planet transits with no evidence for distortions.

We identify a trend in the K2 data that photometric precision depends on the location of a star on the detector. Specifically, for bright stars, photometric precision is highest closer to the center of the K2 field of view, and for faint stars, photometric precision is highest about 3.5 degrees from the center of the field of view. We also investigate the noise power spectra of K2 data and find that they are qualitatively similar to that of \Kepler\ data, with elevated noise levels. The SFF reduction technique improves the quality of K2 data so that much of the science done in the original \Kepler\ mission can continue into its two--wheeled extended mission. 

Finally, we note that data taken while the spacecraft's thrusters are firing is anomalous. In future K2 data releases, it could be useful to have a quality flag indicating when thruster fires take place. 

\acknowledgments
We acknowledge the tremendous effort of the K2 team and Ball Aerospace to make the K2 mission a success. We thank the anonymous reviewer for many insightful comments and suggestions. We thank Tom Barclay for helpful conversations and for providing the precise boresight coordinates for the Kepler Two--Wheel Concept Engineering Test. We thank Eric Agol, Suzanne Aigrain, Juliette Becker, Christian Clanton, Daniel Foreman--Mackey, Heather Knutson, Benjamin Montet, and Jonathan Swift for their helpful comments on the manuscript.  Some/all of the data presented in this paper were obtained from the Mikulski Archive for Space Telescopes (MAST). STScI is operated by the Association of Universities for Research in Astronomy, Inc., under NASA contract NAS5--26555. Support for MAST for non--HST data is provided by the NASA Office of Space Science via grant NNX13AC07G and by other grants and contracts. This paper includes data collected by the \Kepler\ mission. Funding for the \Kepler\ mission is provided by the NASA Science Mission directorate. This research has made use of NASA's Astrophysics Data System and the NASA Exoplanet Archive, which is operated by the California Institute of Technology, under contract with the National Aeronautics and Space Administration under the Exoplanet Exploration Program. A.V. is supported by the NSF Graduate Research Fellowship, Grant No. DGE 1144152. J.A.J is supported by generous grants from the David and Lucile Packard Foundation and the Alfred P. Sloan Foundation.

Facilities: \facility{Kepler}


\clearpage

\end{document}